\documentclass[%
 reprint,
 amsmath,amssymb,
 aps,
]{revtex4-1}

\usepackage{graphicx}% Include figure files
\usepackage{dcolumn}% Align table columns on decimal point
\usepackage{bm}% bold math
\begin{document}

\preprint{APS/123-QED}

\title{Probing strain modulation in a gate-defined one dimensional electron system}% Force line breaks with \\

\author{M.~H.~Fauzi}
%\affiliation{Department of Physics, Tohoku University, Sendai 980-8578, Japan}
\affiliation{Center for Spintronics Research Network, Tohoku University, Sendai 980-8577, Japan}

\author{M. F. Sahdan}
\affiliation{Department of Physics, Tohoku University, Sendai 980-8578, Japan}

\author{M. Takahashi}
\affiliation{Department of Physics, Tohoku University, Sendai 980-8578, Japan}

\author{A. Basak}
\affiliation{Department of Electrical Engineering, Indian Institute of Technology Bombay, Mumbai 400076, India}

\author{K. Sato}
\affiliation{Department of Physics, Tohoku University, Sendai 980-8578, Japan}

\author{K. Nagase}
\affiliation{Department of Physics, Tohoku University, Sendai 980-8578, Japan}

\author{B. Muralidharan}
\affiliation{Department of Electrical Engineering, Indian Institute of Technology Bombay, Mumbai 400076, India}

\author{Y. Hirayama}
\affiliation{Center for Spintronics Research Network, Tohoku University, Sendai 980-8577, Japan}
\affiliation{Department of Physics, Tohoku University, Sendai 980-8578, Japan}
\affiliation{Center for Science and Innovation in Spintronics (Core Research Cluster), Tohoku University, Sendai 980-8577, Japan}

\date{\today}% It is always \today, today,

\begin{abstract}
Gate patterning on semiconductors is routinely used to electrostatically restrict electron movement into reduced dimensions. At cryogenic temperatures, where most studies are carried out, differential thermal contraction between the patterned gate and the semiconductor often lead to an appreciable strain modulation. The impact of such modulated strain to the conductive channel buried in a semiconductor has long been recognized, but measuring its magnitude and variation is rather challenging. Here we present a way to measure that modulation in a gate-defined GaAs-based one-dimensional channel by applying resistively-detected NMR (RDNMR) with in-situ electrons coupled to quadrupole nuclei. The detected strain magnitude, deduced from the quadrupole-split resonance, varies spatially on the order of $10^{-4}$, which is consistent with the predicted variation based on an elastic strain model. We estimate the initial lateral strain $\epsilon_{xx}$ developed at the interface to be about $3.5 \times 10^{-3}$.
\end{abstract}

\pacs{Valid PACS appear here}% PACS, the Physics and Astronomy
\maketitle

%\tableofcontents
In many semiconductor-based quantum systems, electrons are manipulated by applying voltages to the surface metal gates. For example, a combination of nanoscale metal gates and GaAs based two-dimensional systems enables us to realize one-dimensional quantum channel and zero-dimensional quantum dot by depleting electrons under the gates\cite{Beenakker}. These building blocks are integrated into many quantum devices, such as quantum computing/simulating systems based on electron spins\cite{Loss, Hanson, Christopher}. Electron control in these systems is always accompanied by electron position change from the originally two-dimensional sheet. One can expect microscopic strain distribution in such devices because surface metal gate and semiconductor system have different thermal expansion coefficients and complicated nanometer surface gates should produce a complicated strain pattern inside. Such phenomena are common for all semiconductor systems including silicon and other semiconductor groups. However, the strain variation felt by confined electrons has not received much attention up to now partly because a lack of appropriate and precise measurement tool to probe local strain in nanometer scale electron channel. Here, taking GaAs-based quantum-point-contact (QPC)\cite{Wharam, Wees} as a prototypical example, we demonstrate that electrons flowing in the one-dimensional channel feel different strain even in the same device when the channel position is microscopically shifted by changing the gate voltage.

There are a couple of methods to measure spatial strain distribution in materials. Examples include X-ray diffraction\cite{Miao, Robinson}, electron microscopy\cite{Martin, Cooper}, and Raman spectroscopy\cite{Mohiuddin, Huang, Neumann}. Although those techniques are capable of delivering a high-spatial resolution strain profile, they are only sensitive to strain magnitude larger than a factor of $10^{-4}$. Alternative technique such as solid-state NMR could provide an acceptable solution since it has the ability to detect ultra low-level strain variation of less than $10^{-4}$ through nuclear quadrupolar interaction with the electric field gradient (EFG)\cite{Zwanziger, Shulman, Sundfors69}. However, macroscopic samples are needed for the conventional NMR detection technique to work. Furthermore, it is difficult to get information of the electron-existing nanometer scale area inside the semiconductors with these techniques.

To overcome the limitation, the so-called optically-detected (or optically-pumped) NMR with quadrupole nuclei has been developed and exploited intensively to investigate structural information of strained semiconductor nanostructures\cite{Flinn, Guerrier, Chekhovich_NatNano, Wood, Ulhaq, Eickhoff, Kuznetsova, Chekhovich_NatComm, Bulutay, Matthew}. However this technique requires an interrogated structure to be optically accessible, which cannot be easily applied to nanostructure transport devices defined by surface gate metals such as quantum point contacts\cite{Wharam, Wees} or lateral surface superlattices\cite{Davies, Larkin, Long}. To circumvent the difficulties, we utilize a resistively-detected NMR (RDNMR) technique where both nuclear-spin polarization and detection can be realized in the electron channel thanks to the successful RDNMR in QPCs \cite{Gervais_book, Fauzi}.

\begin{figure}[t]
\begin{center}    
\centering
\includegraphics[width=\linewidth]{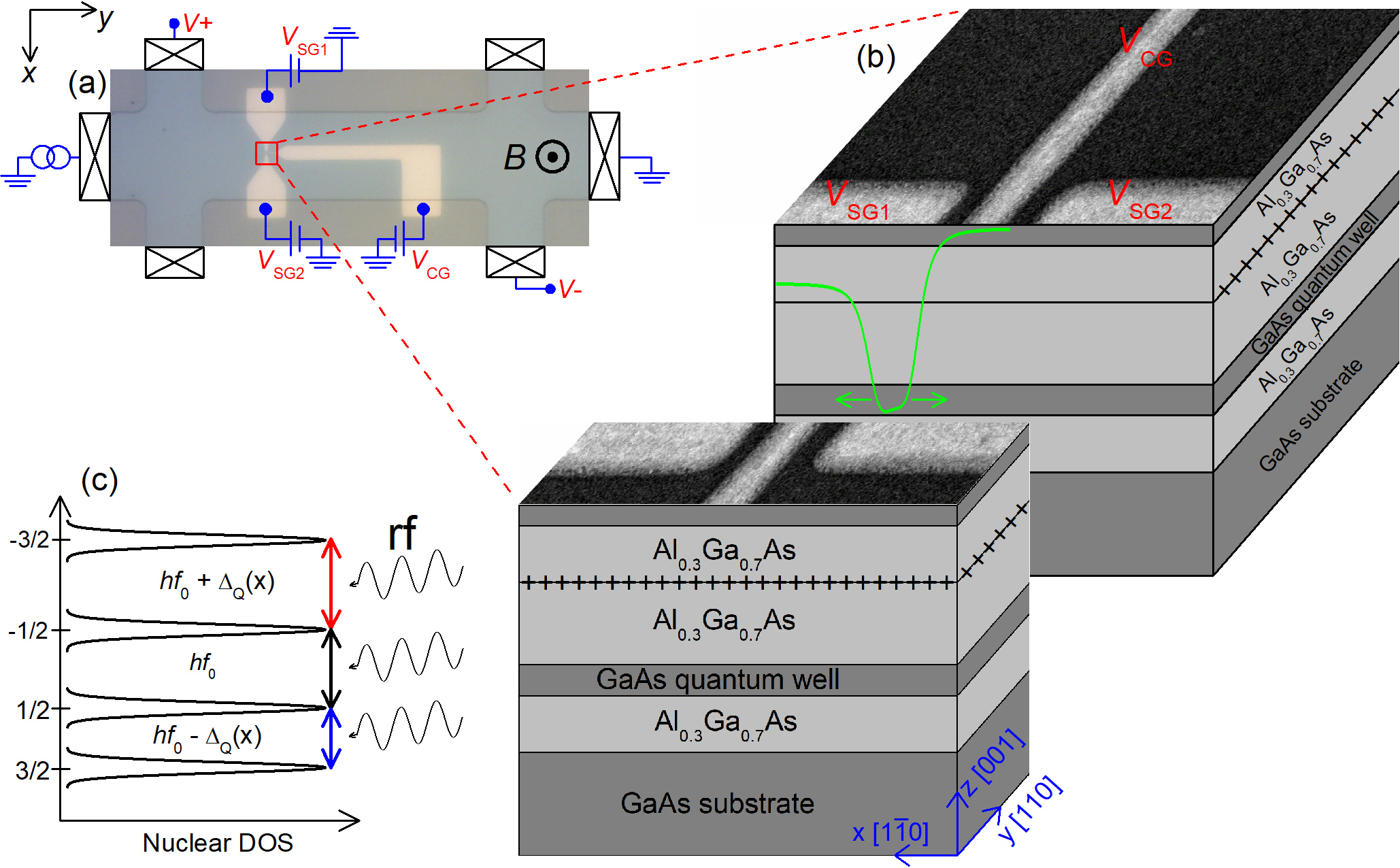}
\end{center}
\caption{(a) An optical image of device layout and transport measurement schematic. (b) A SEM image of the fine metal gates along with a schematic of cut-through wafer structure. Applying negative bias voltages to the three Schottky metal gates ($V_{\rm{SG1}}$, $V_{\rm{SG2}}$, and $V_{\rm{CG}}$) at the surface define the confinement potential (green curve) and control the channel position in lateral direction indicated by the green arrow. The center gate deposited in between a pair of split gates gives us more freedom to tune the confinement potential and thereby allows us to shift the channel over a wider region along x direction than that offered by a traditional point contact device with a pair of split gates only. (c) Anisotropic strained lattices, whose values are positional dependence, create an electric field gradient. An energy level of a 3/2 quadrupole nuclei in close proximity to the field gradient would be affected and can be observed directly through the NMR spectrum. }
\label{Fig01} 
\end{figure}

In this experiment, we used a QPC defined by three independent metallic gates placed on the semiconductor top \cite{Fauzi} as depicted in Fig. 1(a). In RDNMR experiment, we applied perpendicular magnetic field, which pushed the system in quantum Hall regime with edge channels. To avoid possible reflection from the center gate arm connected to the outside of the Hall bar, we fully depleted electron channel between the center gate and split gate 2 by applying a negative bias voltage to $V_{\rm{SG2}}$, which is more negative than a pinch-off bias voltage, naturally depending on a bias applied to the center gate, $V_{\rm{CG}}$ (see the supplementary materials for determining the pinch-off voltage). Figure 1(b) shows a three-dimensional schematic view of electron channel in the QPC. The wafer structure used here puts two-dimensional electron plane at 175 nm from the surface. The quasi-one-dimensional QPC channel is defined by a combination of negative $V_{\rm{SG1}}$ (voltage applied to split gate 1) and $V_{\rm{CG}}$. The channel position can be laterally shifted by tuning $V_{\rm{SG1}}$ and $V_{\rm{CG}}$ as schematically shown in Fig. 1 (b). We start off with the condition where electron channel locates around the edge of the split gate 1. We can expect large strain slope on this situation. Based on many experiments done with different gate voltages, we found that the expected situation can be obtained by applying $V_{\rm{CG}} = -0.45$ V. 

Before going to the detailed experimental results, we will discuss how we obtained RDNMR signal in the electron channel. Dynamic nuclear polarization (DNP) relied on the hyperfine-mediated inter-edge spin flip scattering within the same Landau level as described in our previous theoretical and experimental studies\cite{Fauzi, Aniket}. We applied the magnetic field $4.5$ T perpendicular to the sample to reach the lowest Landau level (filling factor $\nu = 2$) at a lattice temperature of $300$ mK. DNP was induced by applying ac bias current of about $10$ nA for over $1500$ seconds at a certain point along the red conductance traces (see Fig. 2 (a)), corresponding to the filling factor less than 1 ($\nu < 1$) in the constriction. This was followed by slowly scanning rf with increasing frequency through a home-made coils wounded around the device with an rf power of $-30$ dBm delivered to the top of the cryostat and a scanning speed of 100 Hz/s. The QPC conductance is determined by the highest potential at the center of the constriction so that any slight change of the potential height by nuclear Zeeman energy can be sensitively detected in RDNMR. In our previous study in Ref. \cite{Fauzi}, we confirmed that the RDNMR signals were Knight shifted, proving that the detected signals came from inside the constriction where $\nu$ is close to 1. 

As already mentioned, we applied $V_{\rm{CG}} = -0.45$ V to the center metal gate, and then repeated current-induced dynamic nuclear polarization and RDNMR measurements at a certain range of $V_{\rm{SG1}}$ bias voltage along the red line as indicated in the magnetotransport traces displayed in Fig. 2(a). Three represented $^{75}$As RDNMR spectra shown in Fig. 2(b) all exhibit three-fold splitting due to nuclear quadrupole interaction with the strain field. We extracted the average quadrupole splitting value for each obtained RDNMR spectrum with a Gaussian fit. The extracted values are displayed in Fig. 2(c). The detected splitting was initially about 10 kHz at a bias voltage of $V_{\rm{SG1}} = -0.7$ V with the center of each transition peak being slightly convoluted but still recognizable. However, by applying more negative bias voltage to $V_{\rm{SG1}}$, the splitting between the center and satellite peaks progressively increased reaching up to about 25 kHz at $V_{\rm{SG1}} = -1.1$ V. For the case of $V_{\rm{CG}} = -0.45$ V, this increased splitting clearly indicates that electrons in the channel feel different strain when the channel is laterally shifted. Although this result was expected, this experiment is the first to clearly indicate that a slight change in the voltage condition considerably changes the strain in the channel, even within a single QPC device.

\begin{figure}[t]
\begin{center}    
\centering
\includegraphics[width=
\linewidth]{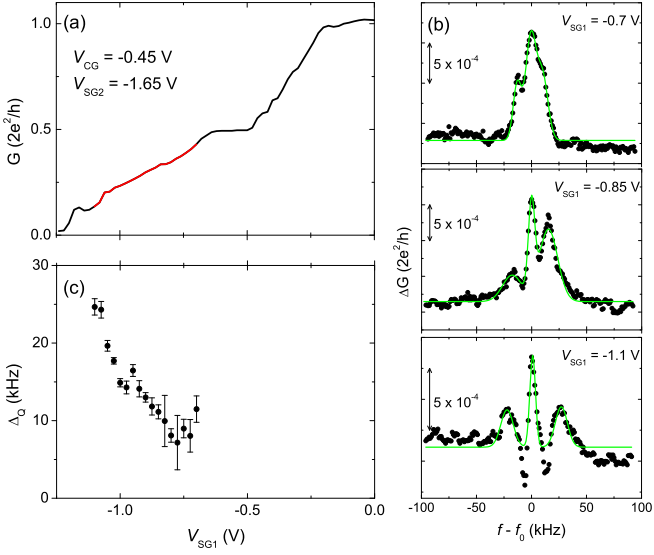}
\end{center}
\caption{(a) Diagonal magneto-transport trace as a function of left-hand side split gate ($V_{\rm{SG1}}$) with $V_{\rm{SG2}} = -1.65$ V and $V_{\rm{CG}} = -0.45$ V. (b) Represented $^{75}$As RDNMR profile where the center to transition peak position increases with applying more negative bias voltage to $V_{\rm{SG1}}$. The solid line is a Gaussian fit to the data where the quadrupole splitting displayed in panel (c) is extracted from. (c) Quadrupole splitting from the spectra measured along the red line.}
\label{Fig02} 
\end{figure}

\begin{figure}[t]
\begin{center}    
\centering
\includegraphics[width=
\linewidth]{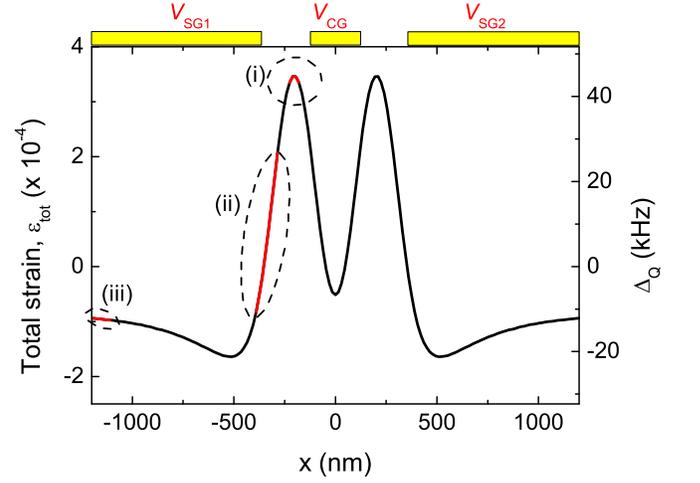}
\end{center}
\caption{Calculated total strain field modulation $\epsilon_{\rm{tot}}$ and corresponding quadrupole splitting $\Delta_Q$ felt by a $^{75}$As nuclei located $175$ nm below the surface. The strain profile has a mirror-symmetry at $x = 0$. Three distinct regions of interest, which are accessible experimentally, are highlighted alphabetically. The total strain reaches a maximum value half-way between the split and center gates, corresponding to region (i). The strain drops rather quickly towards the left split gate (SG1) and changes its sign. The profile is inflected at $x = -500$ nm and the value slowly reduces toward the far left split metal gate (SG1).}
\label{Fig03} 
\end{figure}

\begin{figure}[t]
\begin{center}    
\centering
\includegraphics[width=
\linewidth]{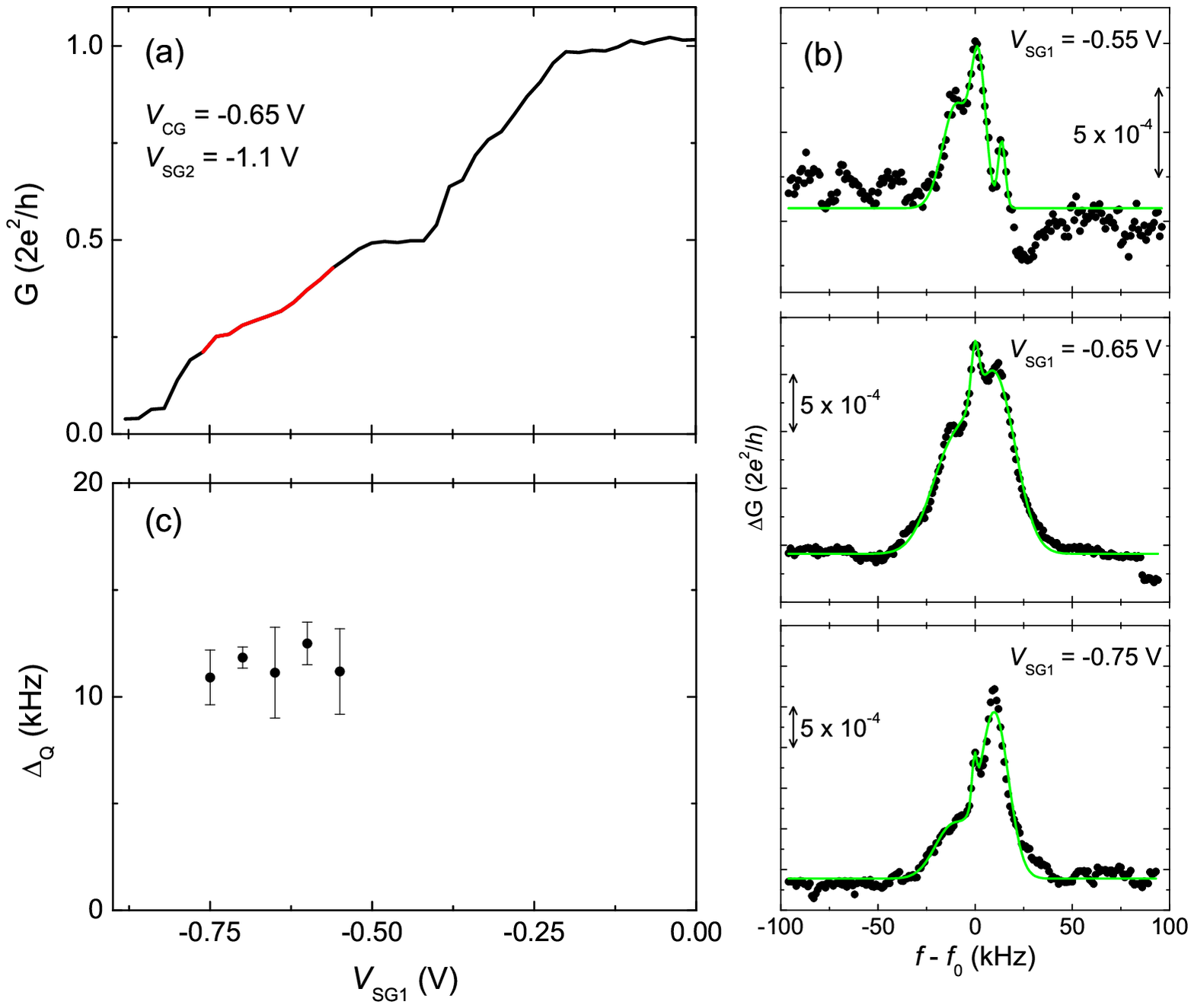}
\end{center}
\caption{(a) Diagonal magneto-transport trace as a function of left-hand side split gate ($V_{\rm{SG1}}$) with $V_{\rm{SG2}} = -1.1$ V and $V_{\rm{CG}} = -0.65$ V. RDNMR spectra were taken along the red line. (b) Represented $^{75}$As RDNMR profile for region (iii). The solid line is a Gaussian fit to the data where the quadrupole splitting displayed in panel (c) is extracted from. (c) Quadrupole splitting from the spectra measured along the red line.}
\label{Fig04} 
\end{figure}

To discuss more quantitatively, we estimate strain distribution in our QPC device with three metallic gates placed on the semiconductor top as depicted in Fig. 1(a). Each metal gate exerts a stress on the semiconductor due to different coefficient of thermal expansion; correspondingly, the resultant of the stressors produces a lateral strain field modulation in the channel. To quantitatively assess the strain profile, we analytically calculate the strain propagation from the interface down to the semiconductor layer on the basis of the model introduced by Davies and Larkin\cite{Davies, Larkin}, as displayed in Fig. 3. The Davies-Larkin model operates under the assumption that there is no displacement in the $y$-direction ($\epsilon_{yy} = 0)$ although its stress component $\sigma_{yy}$ is not zero. In this case, the dilation ($\delta \equiv \epsilon_{xx} + \epsilon_{zz}$) is given by

\begin{equation}
E_{\rm{GaAs}}\delta = (1+\nu_{\rm{GaAs}})(1-2\nu_{\rm{GaAs}})(\sigma_{xx}+\sigma_{zz})
\end{equation}
here $E_{\rm{GaAs}}$ is the Young's modulus of GaAs and $\nu_{\rm{GaAs}}$ is the poisson ratio of GaAs (not to confuse with the filling factor). The stress component on the right-hand side of equation (1) can be computed semi-analytically by taking the real component of the first derivative of the so-called elastic potential shown inside the bracket of equation (2)

\begin{equation}
\sigma = \frac{F_x ({z = 0})}{\pi}\Re\frac{d}{dx}\left (\ln{\frac{\sin{\frac{1}{2}}({Z - A})}{\sin{\frac{1}{2}}({Z + A})}} \right)
\end{equation}
here $Z = \pi(x + iz)/(a + b)$ and $A = \pi a/(a + b)$ describe the gate geometry. The gate length and the gap between each gate are $2a$ and $2b$, respectively. Since the length of the center ($200$ nm) and split metal gate (set to $3$ $\mu$m-long) are different, we compute the strain/stress profile of each gate by taking $b \rightarrow \infty$. The resultant strain/stress is the sum of each individual strain/stress profile. The force per unit length concentrated at the edge $F_x ({z = 0})$ is given by

\begin{equation}
F_x ({z = 0}) = \frac{hE_{\rm{gate}}}{1-\nu_{\rm{gate}}}\epsilon_{xx}({z = 0})
\end{equation}

In our case, the metal gate thickness is approximately $h \approx 25$ nm, which is much smaller than the center or split metal gate length. $E_{\rm{gate}} =  100$ GPa and $E_{\rm{GaAs}} = 85.5$ GPa are the Young's modulus of the gate and GaAs, respectively. The poisson ratio of the gate and GaAs are $\nu_{\rm{gate}} = 0.3$ and $\nu_{\rm{GaAs}} = 0.31$, respectively. The initial differential thermal contraction is $\epsilon_{xx}({z = 0}) = \alpha \times 10^{-3}$. We set the coefficient $\alpha$ to be $3.5$ to match the experimental data. Note that there is uncertainty in the literature about the initial strain/stress coefficient value $\alpha$, therefore we might treat it as a free parameter, and the only free and adjustable parameter in the model calculation. The uncertainty arises from the annealing condition during the gate deposition, which adds extra strain to the interface\cite{Long}. 

Individual strain component $\epsilon_{xx}$ and $\epsilon_{zz}$ are related by $\epsilon_{xx}/{\epsilon_{zz}} = \left({\nu_{\rm{GaAs}} - 1}\right)/{\nu_{\rm{GaAs}}} \approx -2.2258$. Each strain tensor component can be extracted from the computed dilation to evaluate the total strain $\epsilon_{\rm{tot}} = \epsilon_{\rm{zz}} - (\epsilon_{\rm{xx}} + \epsilon_{\rm{yy}})/2$ felt by a nuclei. The corresponding first-order quadrupole splitting $\Delta_Q$ is directly proportional to the strain field $\epsilon_{\rm{tot}}$, which is given by
\begin{equation}
\Delta_Q = \frac{eQS_{11}}{2h}\epsilon_{\rm{tot}}
\end{equation}
here $e$ is the elementary charge and $h$ is the Planck constant. For $^{75}$As nuclei, the EFG tensor component $S_{11} =  \mp13.2 \times 10^{15}$ statC$\cdot$cm$^{-3}$ ($S_{11} = \mp3.96 \times 10^{22}$ Vm$^{-2}$ in SI unit)\cite{Sundfors}, and a quadrupole moment $Q = 2.7 \times 10^{-29}$ m$^{2}$. The relation between $\epsilon_{\rm{tot}}$ and $\Delta_Q$ to the individual strain tensor components ($\epsilon_{\rm{xx}}$, $\epsilon_{\rm{yy}}$, and $\epsilon_{\rm{zz}}$) immediately implies that the quadrupole interaction is only sensitive to shear lattice deformation, but not to isotropic deformation\cite{Chekhovich_NatNano}.

The Davies-Larkin model provides an estimate of the strain field magnitude and its spatial modulation to aid in our discussion. The magnitude varied from about $-1.6 \times 10^{-4}$ to $+3.4 \times 10^{-4}$ at the center of the quantum well, located 175 nm beneath the surface. Since GaAs is in tension, the strain is mostly positive on the exposed surface and takes on a maximum value of $+3.4 \times 10^{-4}$ half-way between the center and split gate. However, the region under the gate has a mostly negative strain field value. The positive(negative) value of $\epsilon_{\rm{tot}}$ means that the crystal lattice in the $x$ direction is subjected to compressive (tensile) strain while the lattice in the $z$ direction is subjected to tensile(compressive) strain. As plotted in Fig. 3, the strain distribution to the left ($x < 0$) and to the right ($x > 0$) side of the center gate is identical. But, we use only left side in our present experiments. Around the edge of the split metal gate shown in (ii) in Fig. 3, the $\Delta_{\rm{Q}}$ ranges from $0$ to $30$ kHz, showing a good consistency with experimental results obtained in Fig. 2.

To further confirm our understanding, next, we set $V_{\rm{CG}} = -0.65$ V. The electron channel pushed far underneath split metal gate 1 in this condition. Current-induced dynamic nuclear polarization and RDNMR measurements were carried out at a certain range of $V_{\rm{SG1}}$ bias voltage along the red line as indicated in the magnetotransport traces displayed in Fig. 4(a). Three represented $^{75}$As RDNMR spectra shown in Fig. 4(b) all exhibit three-fold splitting due to nuclear quadrupole interaction with the strain field, although the splitting is quite small. We extracted the average quadrupole splitting with a Gaussian fit. The extracted values are displayed in Fig. 4 (c). The splitting was consistent around $10$ kHz, unchanged throughout the bias voltage range of interest. This suggests that the nuclear spins were polarized in a small and also less modulated strained region indicated by (iii) in Fig. 3. 

\begin{figure}[t]
\begin{center}    
\centering
\includegraphics[width=
\linewidth]{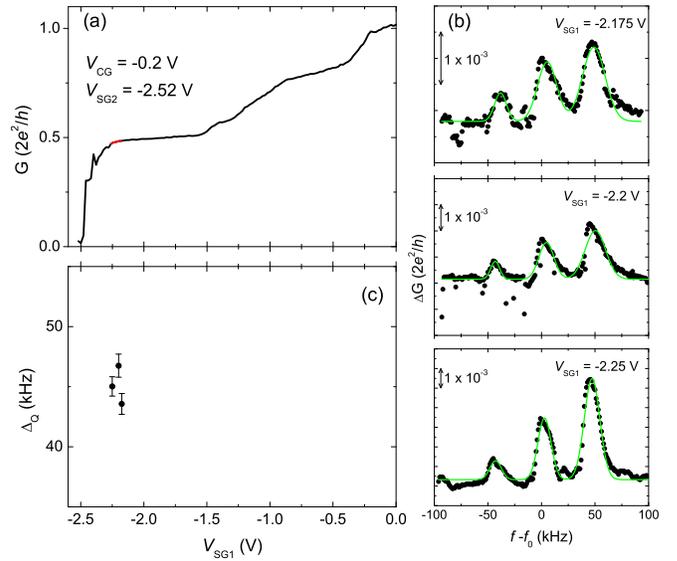}
\end{center}
\caption{(a) Diagonal magneto-transport trace as a function of left-hand side split gate ($V_{\rm{SG1}}$) with $V_{\rm{SG2}} = -2.52$ V and $V_{\rm{CG}} = -0.2$ V. (b) Represented $^{75}$As RDNMR profile half-way between the left split and center gate metal with each peak being clearly separated. (c) Quadrupole splitting from the spectra measured along the red line.}
\label{Fig05} 
\end{figure}

On the other hand, we also try to reduce the applied bias voltage to the center gate to $V_{\rm{CG}} = -0.2$ V to be able to approach the strain field in the exposed area half-way in between the left-hand side split and center metal gates ((i) in Fig. 3), where according to our model, a maximum strain field is expected. Unlike the other two former cases, we notice that the conductance quickly went to zero after passing through the last half-integer plateau as shown in Fig. 5 (a). This occurred because the channel width was already too narrow and consequently we could only accumulate a limited number of spectra to the left vicinity of the plateau, indicated by the red-colored trace. Figure 5 (b) shows the accumulated spectra where each peak was clearly separated since the splitting, of about 45 kHz, has already exceeded the linewidth of each resonance peak. From the splitting value and the channel narrowness, we estimate the nuclear spin polarization detected occupying a volume of around $100 \times 500 \times 20$ nm$^3$, involving about $10^7$ nuclear spins. Since each peak intensity was clearly deconvoluted, the nuclear spin temperature could be estimated easily from the ratio of two satellite intensities \cite{Anant} of around $-2$ mK, indicating that the nuclear spins are population inverted. The detected spectrum was similar to the calculated RDNMR response for relatively large and homogeneously strained $^{75}$As atoms\cite{Aniket}. This is in contrast with the other two former cases where the center transition intensities were mostly found to be more pronounced. Ref. \cite{Matthew} argue that the more pronounced center transition intensity is likely due to the nuclear spin polarization spreads over to the unstrained $^{75}$As atoms. To clearly identify them, it requires a more elaborate 2D strain modelling in combination with self-consistent electron density distribution calculation.

In summary, we have demonstrated direct detection of the built-in strain modulation on the order of $10^{-4}$ in the nanometer-scale channel by electrical means and identified different strain regions. The detection was possible in part since we were able to guide the spin polarized edge current pathways to a different portion of the channel by gate bias tuning. The sensitivity of our strain measurement is currently limited by the center transition linewidth broadening of more than $10$ kHz due to the coupling via inhomogeneous Knight field reflecting electron density distribution in the channel \cite{Chida}. However, it is possible to improve the detection sensitivity by a factor of five at most by depleting the electron density in the channel after each DNP cycle as described in Ref. \cite{Chida, Kumada07}. One can then reduce the central transition linewidth to be as small as $2$ kHz\cite{Chida}, the lower limit due to the nuclear dipolar interaction. 

Evaluation of strain field and its distribution sensed by electrons in a single gate-defined nanostructures is important to understand transport phenomena better in mesoscopic systems as it may alter the confinement potential shape either via deformation potential or piezo-electric coupling\cite{Larkin}. This is particularly relevant for a shallow conductive channel involving multiple gate arrays to study transport anomaly such as the enigmatic 0.7 structures in quantum point contacts, which proved to be sensitive to the confinement potential profile\cite{Burke, FBauer, Iqbal}.

%\acknowledgement

We would like to thank K. Muraki of NTT Basic Research Laboratories for supplying high quality wafers for this study. We thank K. Hashimoto, T. Tomimatsu, and T. Aono for helpful discussions and/or technical assistance. K.N. and Y.H. acknowledge support from the Graduate Program in Spintronics, Tohoku University. Y.H. acknowledge financial support from KAKENHI Grants Nos. 18H01811 and 15H05867. MHF  acknowledge financial support from KAKENHI Grant No. 17H02728.

%\bibliography{ref}

\end{document}